\documentclass{optica-article}

\pdfoutput=1
\journal{optcon}

\articletype{Research Article}
\usepackage{subcaption}	
\usepackage{lineno}

\begin{document}

\title{Fiber-coupled plug-and-play heralded single photon source based on $\mathrm{Ti {:} LiNbO_3}$ and polymer technology}

\author{Christian Kießler \authormark{1,*}, Hauke Conradi\authormark{2}, Moritz Kleinert \authormark{2}, Viktor Quiring \authormark{1}, Harald Herrmann \authormark{1}, and Christine Silberhorn \authormark{1}}

\address{\authormark{1}Integrated Quantum Optics, Institute for Photonic Quantum
Systems (PHOQS), Paderborn University, Warburger Straße 100, Paderborn, 33098, Germany\\
\authormark{2}Fraunhofer HHI, Einsteinufer 37, Berlin, 10587, Germany}

\email{\authormark{*}christian.kiessler@uni-paderborn.de} 



\begin{abstract}
A reliable, but cost-effective generation of single-photon states is key for practical quantum communication systems. For real-world deployment, waveguide sources offer optimum compatibility with fiber networks and can be embedded in hybrid integrated modules. Here, we present the first chip-size fully integrated fiber-coupled Heralded Single Photon Source (HSPS) module based on a hybrid integration of a nonlinear lithium niobate waveguide into a polymer board. Photon pairs at $810 \, \mathrm{nm}$ (signal) and $1550 \, \mathrm{nm}$ (idler) are generated via parametric down-conversion pumped at $532 \, \mathrm{nm}$ in the $\mathrm{LiNbO_3}$ waveguide . The pairs are splitted in the polymer board and routed to separate output ports. The module has a size of $(2 \times 1)\, \mathrm{cm^2}$ and is fully fiber-coupled with one pump input fiber and two output fibers. 
We measure a heralded second-order correlation function of $g_h^{(2)}=0.05$  with a heralding efficiency of $\eta_h=4.5\, \mathrm{\%}$ at low pump powers.
\end{abstract}

\section{Introduction}\label{sec1}
Single photons are suitable as information carriers in quantum computation and communication \cite{Obrien2008,Krenn2016}. One of the the most important building blocks is an efficient source for the generation of single photons. Different approaches are studied to build a practicable single photon source, such as quantum dots \cite{Arakawa2020,Senellart2021,Senellart2017} or nonlinear photon-pair generation with heralding schemes \cite{Krapick2013,Shukhin2020,Paesani2020,Bock2016,Tanzilli2001,Fujii2022,Meany2014}. Efficient plug and play sources have already been demonstrated, being fully integrated with fiber-coupled outputs \cite{Montaut2017,Schlehahn2018}. A major challenge for these sources is an efficient coupling of the generated photons to optical fibers, which is necessary to use these sources practicable. There are commercially available single photon sources based on quantum dot technology, which can be supplied with the associated control electronics and helium cooling \cite{Vergyris2022}. Nevertheless, these sources are very large and expensive. Heralded sources do not need such complex cooling schemes and are thus better suited for further miniaturization. First commercial  sources are available as bench-top devices \cite{Aurea2023}, but they are still far away from the ultimate goal of realizing a quantum- system-on-chip (QSoC) providing low cost, small size and robustness, which are the essential features for future commercial exploitation. \\

\begin{figure}[ht]
	\centering\includegraphics[width=1.0\textwidth]{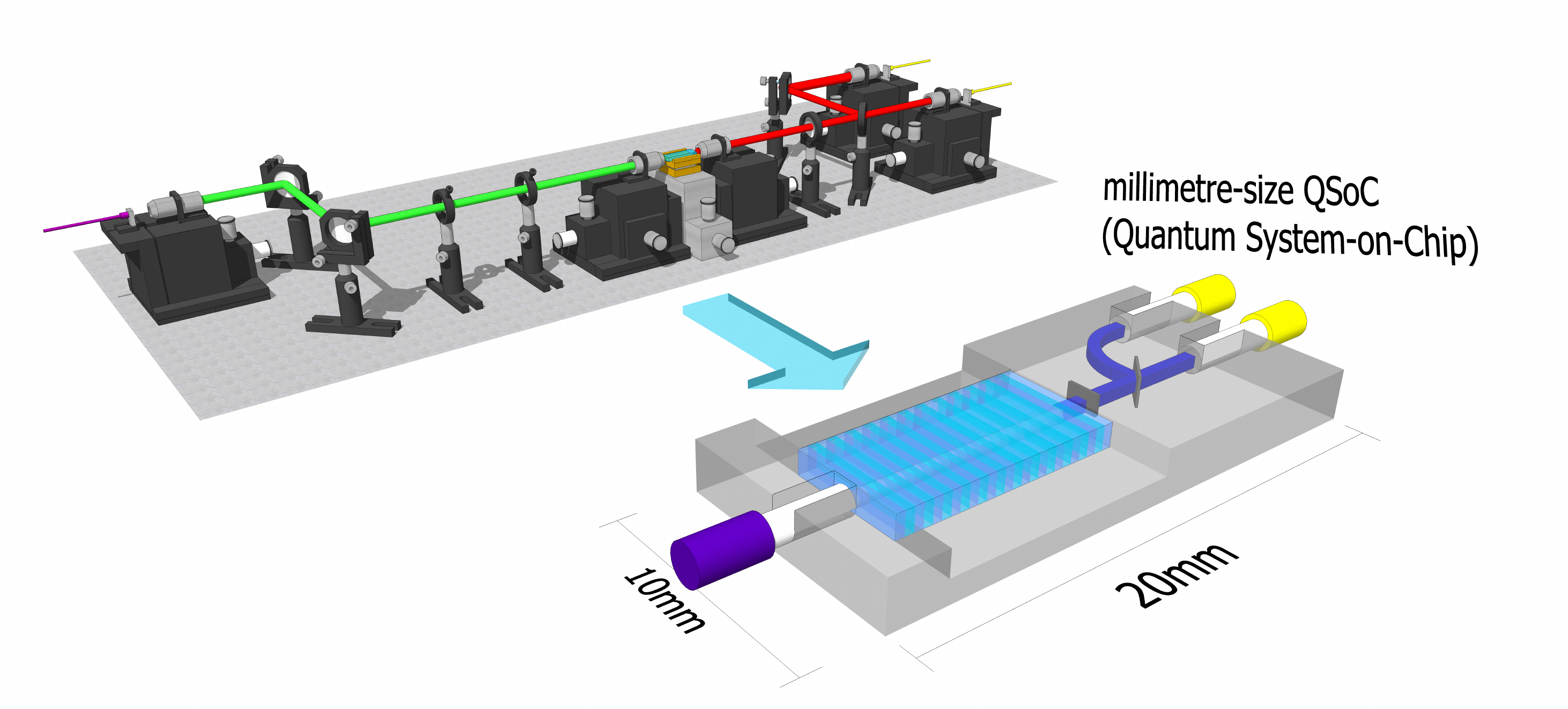}
	\caption{Comparison of a practicable HSPS breadboard setup based on a nonlinear crystal and freespace optics and the corresponding QSoC.}
	\label{Fig1}
\end{figure}

To understand the process and required components of a HSPS, a practicable breadboard setup of a HSPS based on a nonlinear crystal and free-space optics is shown in Figure \ref{Fig1} (top). The pump field is coupled into the nonlinear crystal to generate photon pairs, which are then collected by an out-coupling lens. A filter to suppress the transmitted pump wave and an optical element for separating signal and idler are needed to complete the setup. Due to the probabilistic nature of the nonlinear generation process, a heralding scheme needs to be implemented, which means one photon of the created pair has to be detected to herald the other photon. This gives information on whether there is a single photon in the output or not. The most important parts of a HSPS are therefore the photon-pair generation, pump filtering, signal/idler separation and the detection. Ultimately, in a fully integrated source all of these components must be combined in one module, which has not been demonstrated yet. Sufficient pump filtering is difficult to implement on chip in an integrated device. Various non-integrated approaches to filter out pump light by using fiber-coupled filters placed after the module have been presented \cite{Montaut2017}. However, for a compact fully integrated source, this filtering must be done on chip. The detection of one of the photons should also be directly integrated in the module. An approach to combine everything in one device is the hybrid integration of multiple materials with different optical properties and fabrication capabilities. 
\\
In this paper, we present a fully integrated chip-size, plug-and-play HSPS module based on lithium niobate and polymer technology. A polymer board (PolyBoard) is used as a chip platform, which enables the fabrication of waveguides and etched slots for the integration of different optical components \cite{Keil2012,Kleinert2017}. Filters with various spectral and polarization characteristics can be glued into these etched slots, which open up the possibility for sufficient pump filtering and signal/idler separation. The PolyBoard is therefore a well suited platform for the integration of different optical components on chip. A SPDC process in a titanium-indiffused periodically poled z-cut lithium niobate waveguide ($\mathrm{Ti {:} LiNbO_3}$) is designed to generate photon pairs at $810\, \mathrm{nm}$ (signal) and $1550\, \mathrm{nm}$ (idler). The high second order nonlinearity of $\mathrm{LiNbO_3}$ makes it a favorable choice for an efficient SPDC process \cite{Weis1985}. The signal photons are used to herald the telecom photons. The $\mathrm{Ti {:} LiNbO_3}$ waveguide is butt-coupled to the PolyBoard waveguide. The hybrid integration of a $\mathrm{Ti {:} LiNbO_3}$ waveguide into the PolyBoard leads to a fully integrated module, which cannot be realized in one material alone. The signal wavelength of $810\, \mathrm{nm}$ enables the possibility for a future integration of silicon single-photon avalanche diodes (SPADs) to the module and therefore a possible commercially available device with small size, low cost, high robustness and reproducibility. \\
A model of the QSoC HSPS is shown in Figure \ref{Fig1} (bottom). In the following we describe the general design and pre-characterization of single components of the module. An analysis of the single photon characteristics follows afterwards.

\section{Design and components}\label{sec2}
We use the PolyBoard technology as the QSoC platform for the HSPS-module. The general design of the module is described in chapter \ref{subsec2}. We integrate a periodically poled $\mathrm{Ti {:} LiNbO_3}$ waveguide as nonlinear medium for the SPDC process into the PolyBoard. The design and characterization of the $\mathrm{Ti {:} LiNbO_3}$ is presented in the following. 

\subsection{Periodically poled $\mathrm{Ti {:} LiNbO_3}$}\label{subsec1}
\begin{figure}[ht]
	\centering
	\begin{subfigure}[c]{0.5\linewidth}
		\centering
        \subcaption{}
		\includegraphics[width=\linewidth]{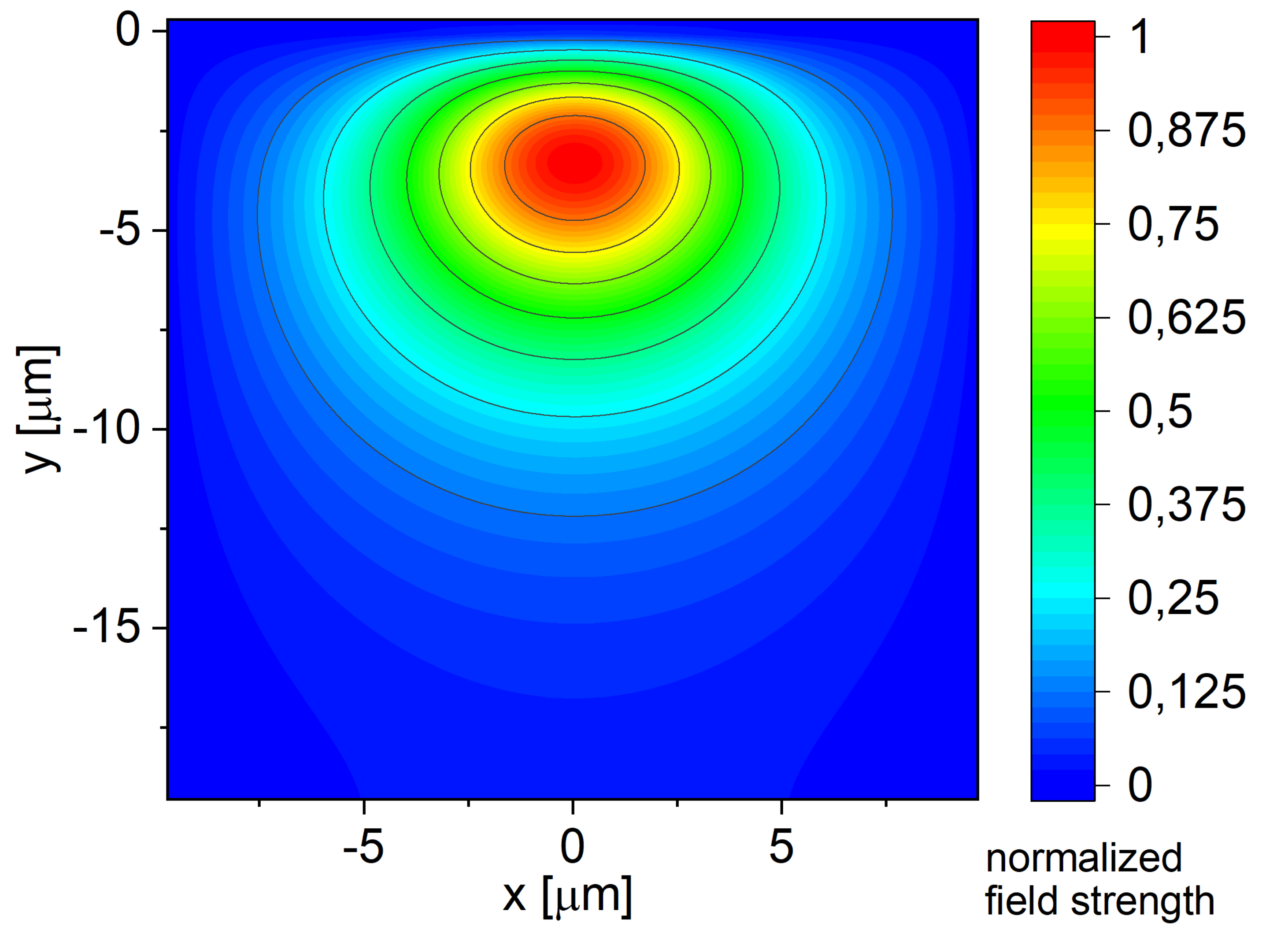}
		\label{Fig5}
        	\end{subfigure}
	\begin{subfigure}[c]{0.45\linewidth}
		\centering
        \subcaption{}
		\begin{tabular}{c|c|c} 
			$\lambda$ & LiNbO$_3$ & PolyBoard \\ \hline
			$532\,\mathrm{nm}$ &  $3.8\, \mathrm{\mu m} \times 2.6\, \mathrm{\mu m}$ &  - \\
			$810\,\mathrm{nm}$ &  $5.7\, \mathrm{\mu m} \times 3.9\, \mathrm{\mu m}$ & $3.4\, \mathrm{\mu m}$ \\ 
			$1550\,\mathrm{nm}$ &  $8.9\, \mathrm{\mu m} \times 6.8\, \mathrm{\mu m}$ & $6.4\, \mathrm{\mu m}$\\ 		
		\end{tabular}
		\label{Tab1}
       	\end{subfigure}
    \caption{Simulated $\mathrm{Ti {:} LiNbO_3}$ mode field distribution of the $1550\, \mathrm{nm}$ idler TM-mode (a); MFDs at different wavelengths $\lambda$ for $\mathrm{Ti {:} LiNbO_3}$ and PolyBoard waveguides (b).}
\end{figure}

\begin{figure}[ht]
    \centering
	\includegraphics[width=0.75\linewidth]{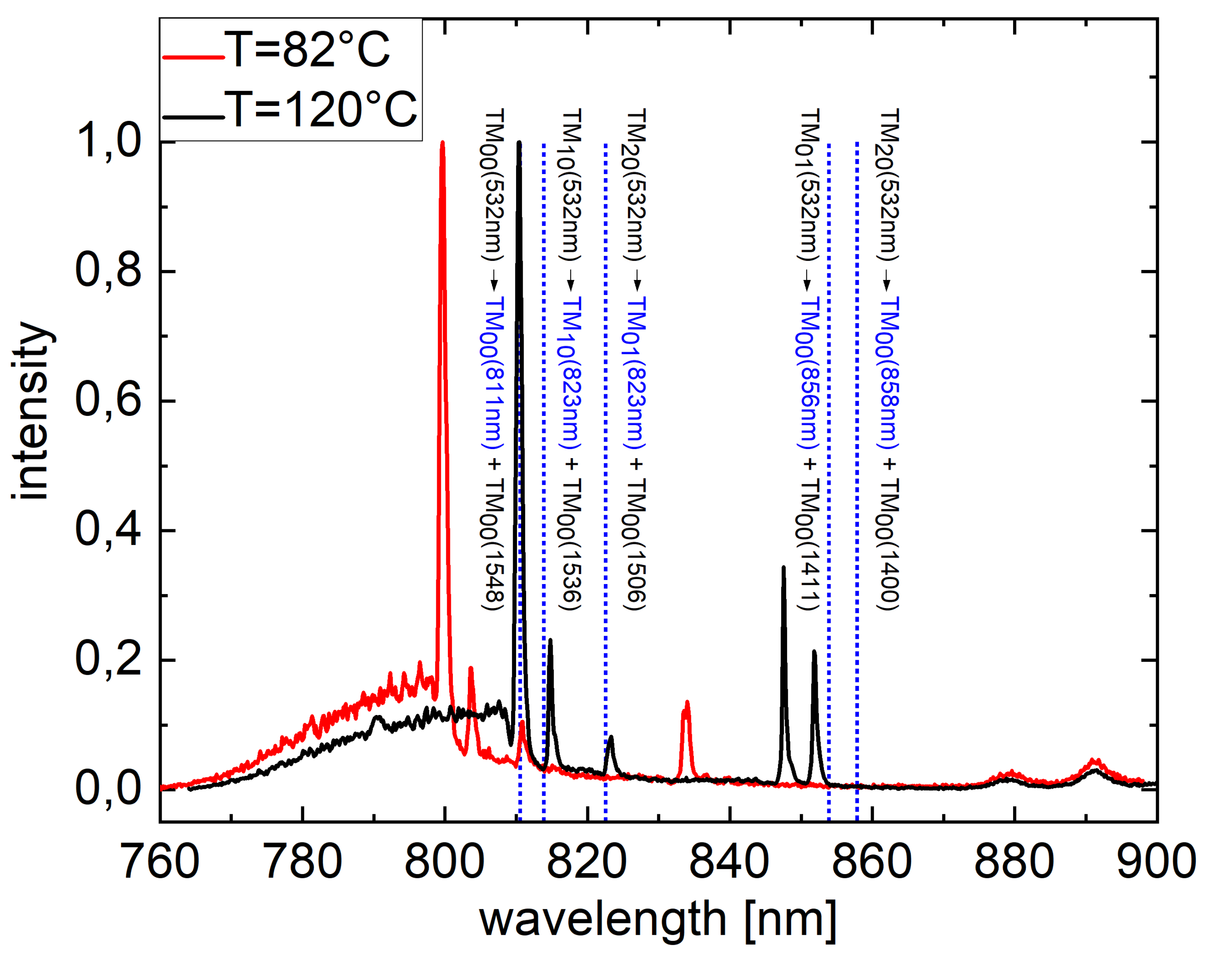}
	\caption{Measured SPDC spectrum of the periodically poled $\mathrm{Ti {:} LiNbO_3}$ for a poling period of $7.05 \, \mathrm{\mu m}$ at different temperatures. Calculated phasematching peak positions for different mode combinations are marked in blue.}
	\label{Fig2}
\end{figure}

\begin{figure}[ht]
    \centering
	\includegraphics[width=0.75\linewidth]{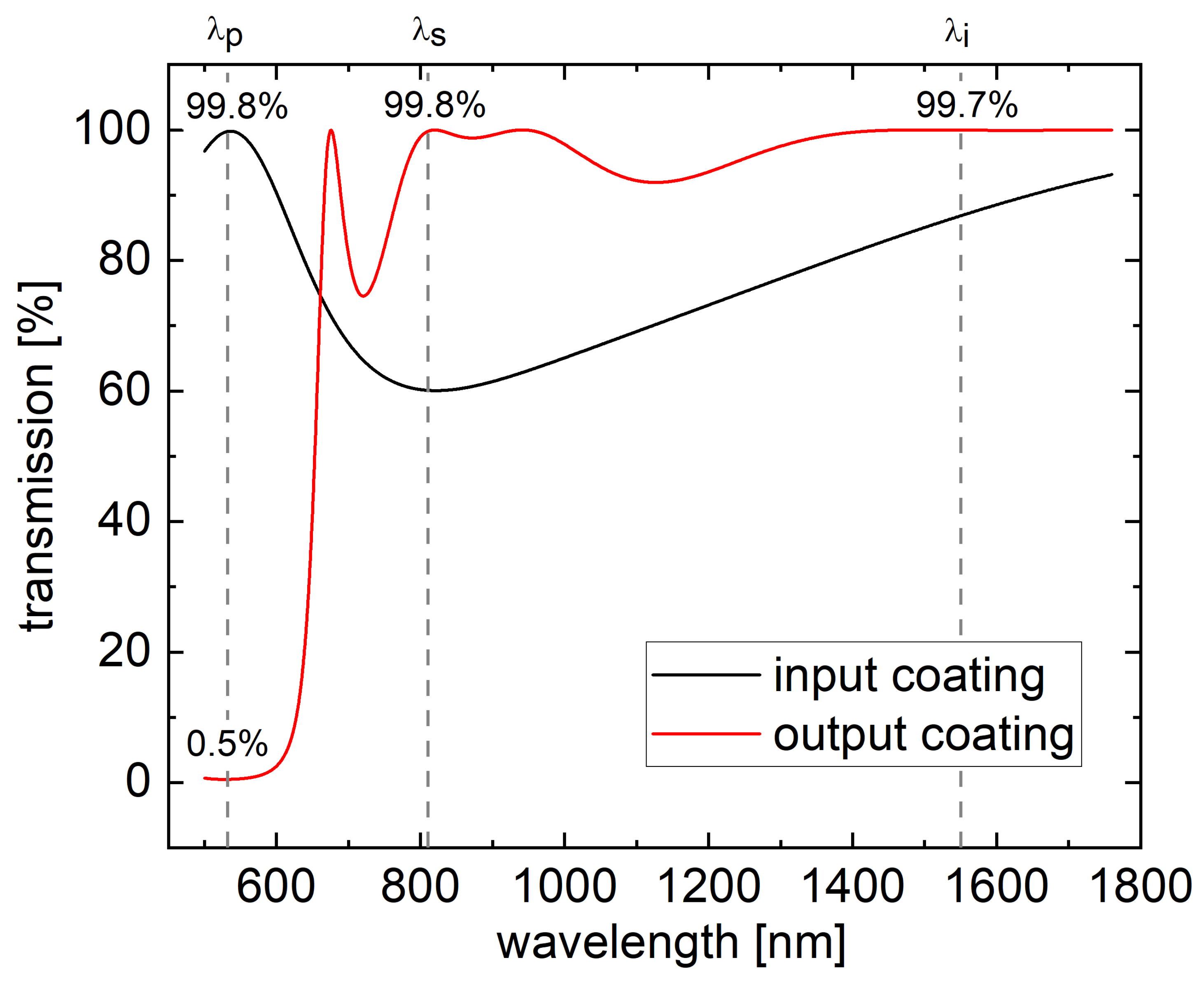}   
    \caption{Measured and converted transmission spectra of the $\mathrm{Ti {:} LiNbO_3}$ in- and output coatings.}
    \label{Fig3}
\end{figure}

We use z-cut $\mathrm{LiNbO_3}$ as the nonlinear material for the desired SPDC process, because of its high second order nonlinearity and wide transparency window \cite{Weis1985}. Additionally the spontaneous polarization of $\mathrm{LiNbO_3}$ enables a periodic domain inversion of the crystal, which allows for quasi-phase-matched nonlinear processes. We fabricated low loss waveguides (typically $<0.1\, \mathrm{ dB/cm}$) by an indiffusion of $5\, \mathrm{\mu m}$ wide, $80\, \mathrm{nm}$ thick titanium-stripes into the substrate for $8.5\, \mathrm{h}$ at $1060\, \mathrm{^\circ C}$. Periodic poling of the waveguides ensures the quasi-phase-matching condition for the designed type-0 SPDC process with a pump wavelength at $532 \, \mathrm{nm}$ and signal and idler at $810 \, \mathrm{nm}$ and $1550 \, \mathrm{nm}$, respectively . We fabricated several waveguides with various poling periods from $6.80 \, \mathrm{\mu m}$ to $7.15 \, \mathrm{\mu m}$ on one sample. \\ 
For optimal light coupling between different waveguide structures, the mode field distributions must match as best as possible. The simulated mode field distribution of the $\mathrm{Ti {:} LiNbO_3}$ idler TM-mode is shown in Figure \ref{Fig5} examplary. The Ti-indiffusion leads to an asymmetric mode distribution in vertical direction \cite{Minakata1978,Fukuma1978}. At the signal and pump wavelength, the waveguides guide several modes. The corresponding mode field diameters (MFDs, 1/e²) of the fundamental modes at different wavelengths are shown in Table \ref{Tab1}. \\
\\
Before the integration of a $\mathrm{Ti {:} LiNbO_3}$ waveguide into the PolyBoard platform, we pre-characterized the waveguides by analyzing the SPDC process to choose the best working one for the integration. We pumped the sample at $532 \, \mathrm{nm}$ (Klastech cw-laser, $100 \, \mathrm{mW}$) and analyzed the SPDC output  with a spectrometer at visible wavelengths. We controlled the input polarization and power with a half-wave plate and a linear polarizer and used lenses for the waveguide in- and out-coupling. The measured spectrum for a poling period of $7.05 \, \mathrm{\mu m}$ at two different temperatures is shown in Figure \ref{Fig2}.
\newpage
The intensity peak of the generated signal photons at $810\, \mathrm{nm}$ is clearly visible. Smaller peaks at higher wavelengths are due to higher-order mode combinations. We designed the waveguide to guide only a single mode at $1550 \, \mathrm{nm}$, but multiple modes are guided at the other wavelengths. Different mode combinations fullfill the phasematching condition at different wavelengths, which leads to several phasematching peaks in the outptut spectrum. The calculated peak positions for different mode combinations are shown in Figure \ref{Fig2} (blue). The theoretical position of the peak resulting from only fundamental modes is shifted by an offset so that it matches the measured peak. All other calculated values are shifted by this offset, so the distance between different calculated peaks is unchanged. The offset between calculated and measured values is mainly due to fabrication imperfections. A broad structure in the lower wavelength region next to the PDC signal can be identified as nonlinear Cerenkov radiation \cite{Zhang2008}. A higher operation temperature of the $\mathrm{Ti {:} LiNbO_3}$ waveguide not only shifts the PDC signal, but leads to a more stable nonlinear process, which is why temperatures above $90\, \mathrm{^\circ C}$ are optimal. The strong photorefractive effect in $\mathrm{LiNbO_3}$ can lead to mode hopping, which can be suppressed by high operation temperatures \cite{Mondain2020}. Since one $\mathrm{Ti {:} LiNbO_3}$ waveguide is later inserted into a PolyBoard (see \ref{subsec2}), a high temperature would lead to misalignment and damage of the assembled module. Tests have shown, that temperatures above $100\, \mathrm{^\circ C}$ can lead to significant misalignment of the assembled module. Therefore, a compromise regarding the operation temperature has to be found. We have chosen a waveguide with a poling period of $7.05 \, \mathrm{\mu m}$ for the integration into the PolyBoard module, because of a desired temperature of around $80\, \mathrm{^\circ C}$. After characterization, we cut and polished the sample to a final length of $15 \, \mathrm{mm}$ and width of $5 \, \mathrm{mm}$. 
To ensure high transmission between different components, we deposited a dielectric coating consisting of alternating $\mathrm{TiO_2}$ and $\mathrm{SiO_2}$ layers on the endfacets. We designed the output coating to also filter out some of the pump light, since a fully integrated module does not allow for a free-space filtering afterwards. Figure \ref{Fig3} shows the measured spectrum of the input coating with a high transmission at the pump wavelength $T_{in}(532\, \mathrm{nm})=99.8 \, \%$ and the spectrum of the output coating with low transmission for the pump $T_{out}(532\, \mathrm{nm})=0.5 \, \%$ and high transmission for signal $T_{out}(810\, \mathrm{nm})=99.8 \, \%$ and idler $T_{out}(1550\, \mathrm{nm})=99.7 \, \%$. We measured the spectra on a glass substrate and converted them for a transmission between $\mathrm{Ti {:} LiNbO_3}$ and PolyBoard. In addition, we deposited a $200 \, \mathrm{nm}$ protective $\mathrm{SiO_2}$ layer on the top side of the crystal.     
\subsection{PolyBoard}\label{subsec2}
\begin{figure}[ht]
	\centering
	\begin{subfigure}[c]{0.55\linewidth}
		\centering
        \subcaption{}
		\includegraphics[width=\linewidth]{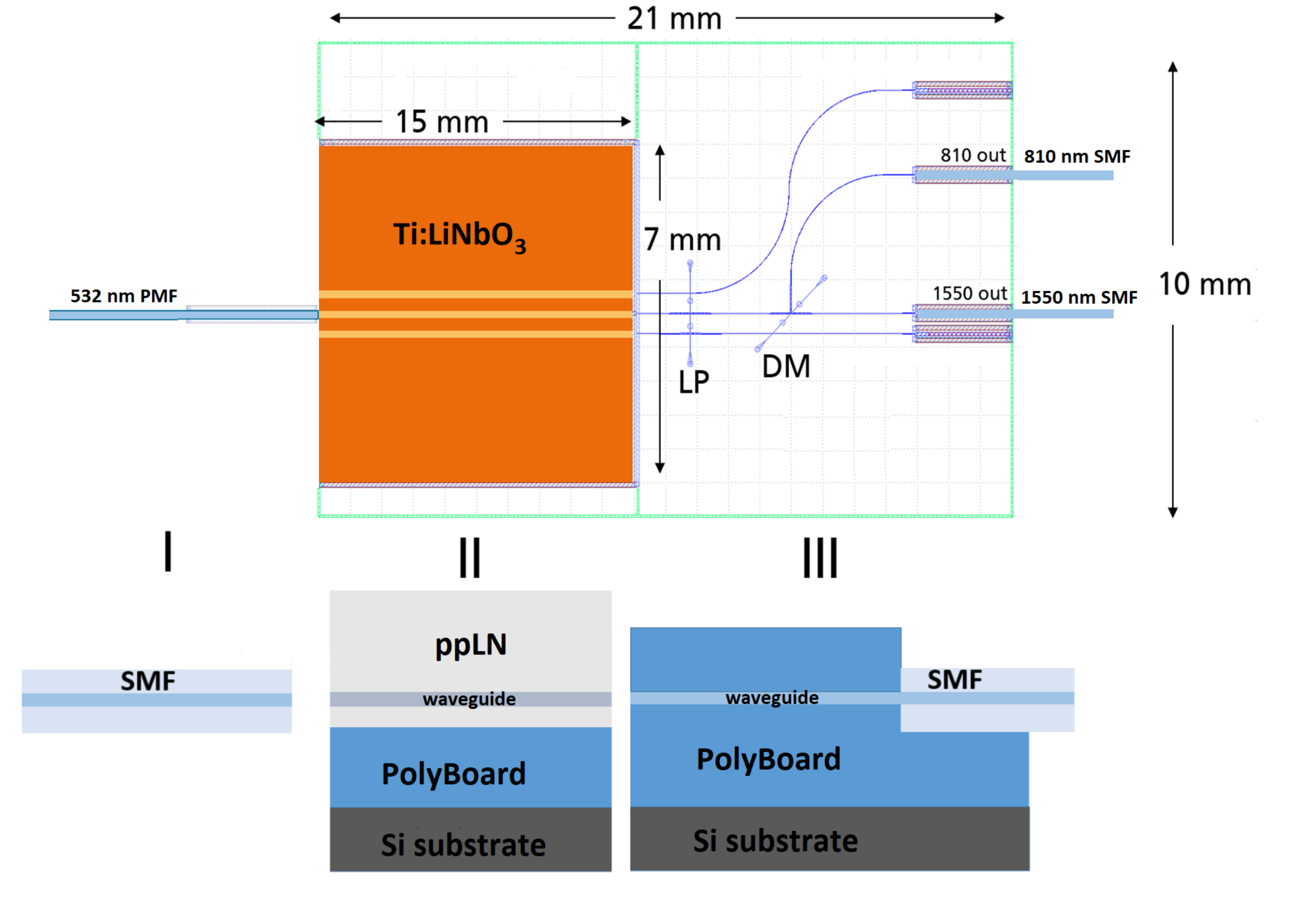}
		\label{Fig6}
    \end{subfigure}
	\begin{subfigure}[c]{0.42\linewidth}
		\centering
        \subcaption{}
		\includegraphics[width=\linewidth]{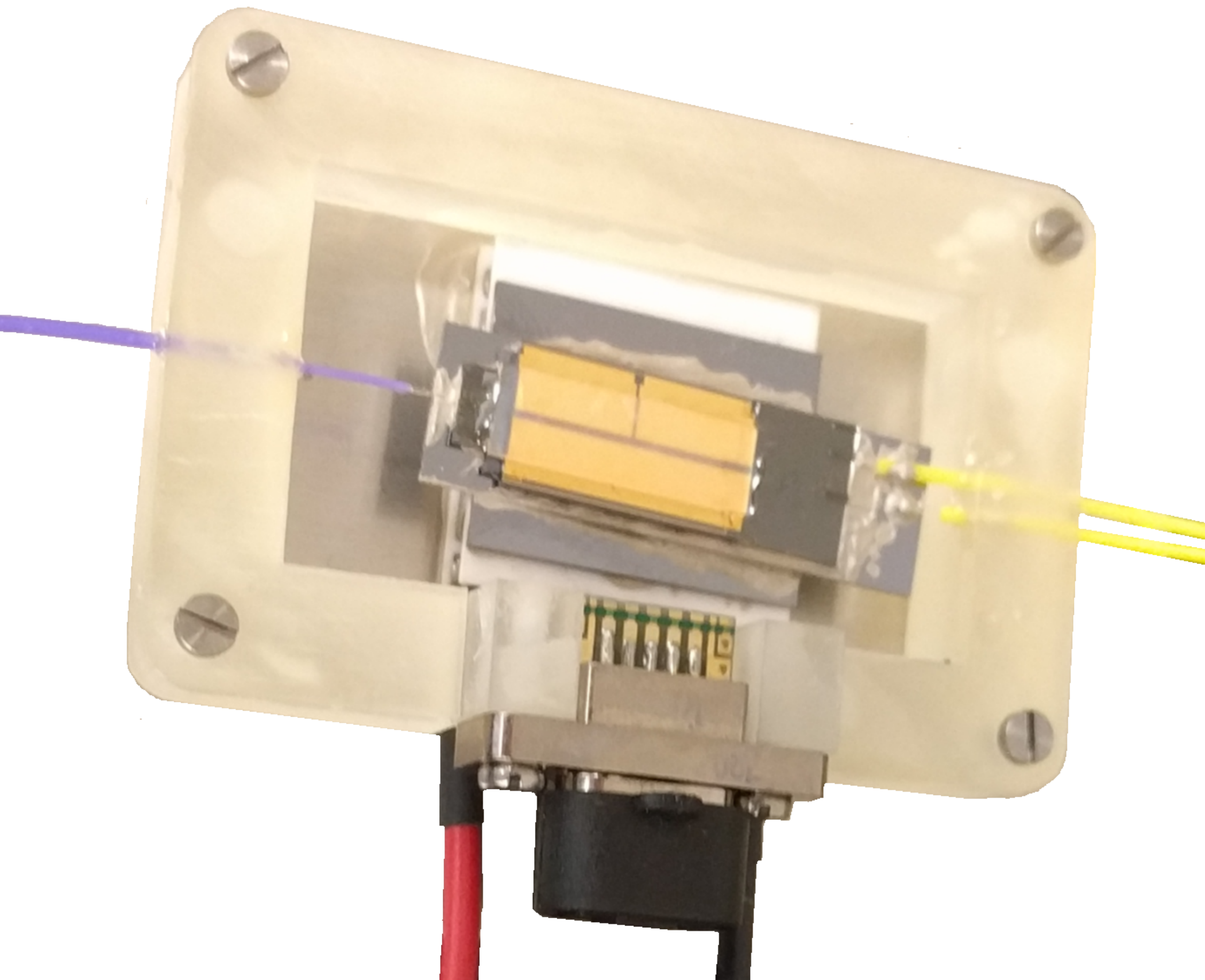}
		\label{Fig7}
    \end{subfigure}
    \caption{Design of the HSPS module (a, top) consisting of $\mathrm{Ti {:} LiNbO_3}$ crystal and PolyBoard with integrated wavguides, long pass filter (LP), dichroic mirror (DM) and side view (a, bottom); assembled and packaged module with PM-input-fiber (purple), two SM-ouput-fibers (yellow) and electrical contacts for temperature control (b).}
\end{figure}

The hybrid integration platform of the HSPS is a perfluorinated acrylate based polymer (ZPU12 series by ChemOptics), which is spin-coated and UV-cured on top of a Si substrate. Square-shaped weakly guiding single mode channel waveguides can be embedded in the PolyBoard. The waveguides are structured by reactive ion etching and the core polymer layers are spin-coated. The square shape ensures a negligible polarization dependence of the waveguides. In addition, U-grooves and slots can be etched for the coupling to optical fibers and the integration of optical elements like thin-film filters (TFFs) into the PolyBoard \cite{Kleinert2019}. The TFFs enable various spectral and polarization filter characteristics and are based on dielectric layer stacks. All this makes the PolyBoard a perfect integration platform for hybrid devices and QSoCs. The biggest challenge of fabricating a fully integrated HSPS is the on-chip pump suppression. The integration of TFFs can solve this problem and is easily expandable. The HSPS module can be divided in three parts (Figure \ref{Fig6}(bottom)), which are glued together. We achieve optimal positioning by maximizing optical transmission through the complete module. We glued the $\mathrm{Ti {:} LiNbO_3}$ on top of a PolyBoard section to increase stability and coupled the input facet of the $\mathrm{Ti {:} LiNbO_3}$ waveguide to a $532 \, \mathrm{\mu m}$ polarization maintaining (PM) optical fiber, since the SPDC process is designed for a TM input polarization, which can be guaranteed by the PM fiber. We coupled the output facet of the $\mathrm{Ti {:} LiNbO_3}$ waveguide to a second PolyBoard with integrated optical components. The design of the complete module is shown in Figure \ref{Fig6} (top).  
\newpage
We implemented three square-shaped weakly guided channel waveguides with an refractive index change of around $\Delta n=0.01$ between core and cladding material \cite{Kleinert2019}. The corresponding MFDs are shown in Table \ref{Tab1}. The PolyBoard waveguides are coupled to $\mathrm{Ti {:} LiNbO_3}$ waveguides, with the center waveguide is coupled to the $\mathrm{Ti {:} LiNbO_3}$ center weaveguide with a poling period of $7.05 \, \mathrm{\mu m}$ (see \ref{subsec1}) and the outer ones only being used for alignment. PolyBoard waveguide losses can be $<1\, \mathrm{dB/cm}$ \cite{Kleinert2019}. We estimate a theoretical coupling efficiency between PolyBaord and $\mathrm{Ti {:} LiNbO_3}$ waveguides of $95 \, \mathrm{\%} \ @ \ 1550\, \mathrm{nm}$ and $87 \, \mathrm{\%} \ @ \ 810\, \mathrm{nm}$ by calculating the overlap integral of the PolyBoard and  $\mathrm{Ti {:} LiNbO_3}$ mode intensity distributions. \\

We etched two deep slots for the insertion of the TFFs into the PolyBoard, with the first TFF implemented as a long-pass filter (LP) for pump suppression and the second one as a dichroic mirror (DM) for separation of signal and idler. The TTFs are inserted by hand into the narrow etched slots and glued afterwards. We determine additional losses due to one TFF insertion of around $1\, \mathrm{dB}$ for both wavelengths. Passive coupling to single mode fibers (SMF) is done via etched U-grooves with fiber-coupling loss of around $1\, \mathrm{dB} \ @ \ 1550\, \mathrm{nm}$ mainly due to MFD mismatch \cite{Kleinert2019}. We estimate an overall transmission loss through the complete module for perfect coupling and alignment by summing up all single losses of $3.1 \, \mathrm{dB} \ @ \ 1550\, \mathrm{nm}$ and $4.2 \, \mathrm{dB} \ @ \ 810\, \mathrm{nm}$. It should be noted that this value is only a lower limit for the overall losses. To control the temperature, we placed a peltier element and a thermosensor below and above the module respectively. We were able to heat the module up to $100 \, \mathrm{^\circ C}$, which is sufficient for operating the HSPS. Higher temperatures should be possible, but are not tested within this work. Figure \ref{Fig7} shows the assembled and packaged HSPS module with PM-input-fiber (purple), two SM-output-fibers (yellow) and electrical contacts for temperature control.  \\

\begin{figure}[ht]
	\centering
	\includegraphics[width=0.6\textwidth]{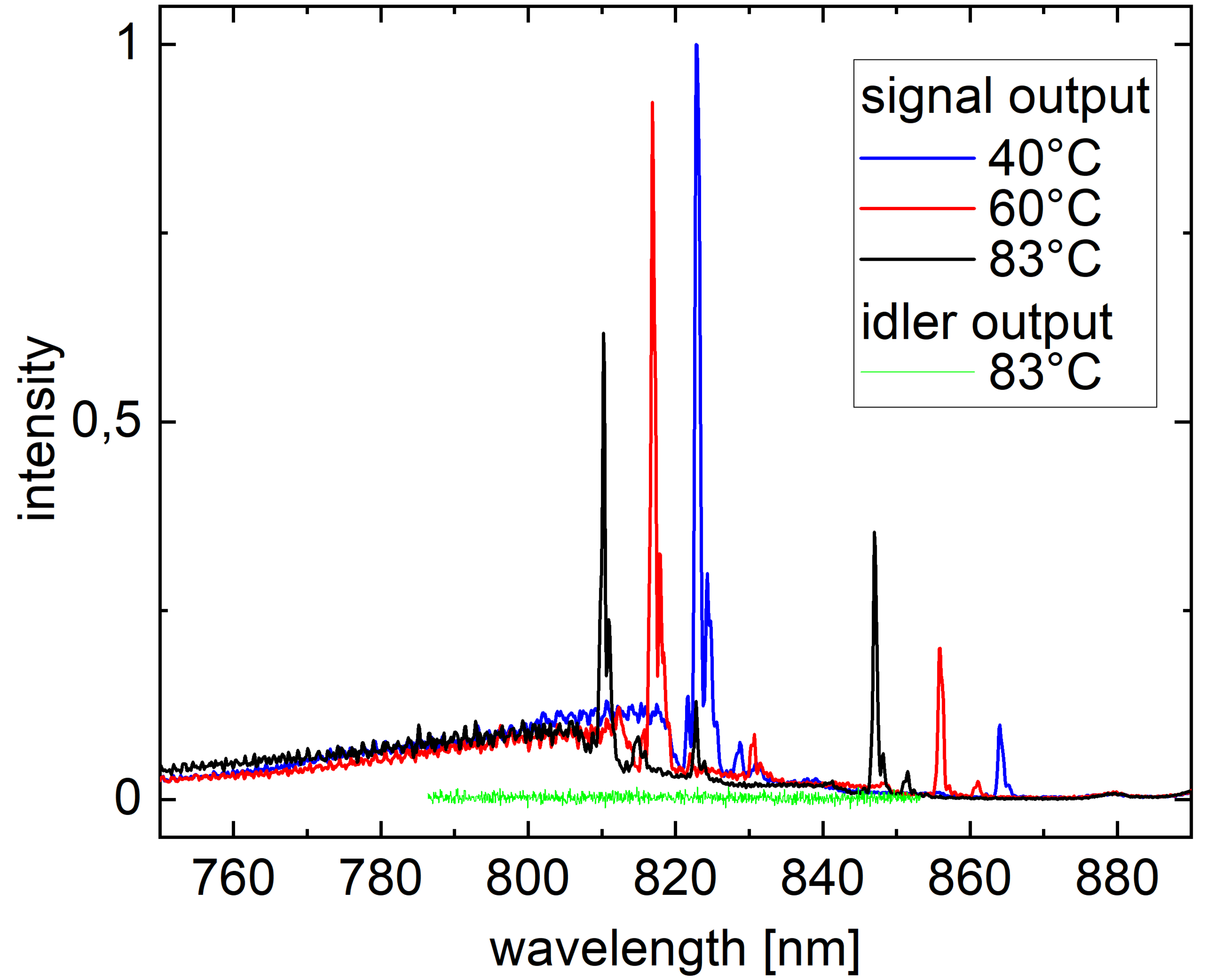}
	\caption{Measured HSPS module output spectra of the signal and idler photons at different temperatures. No counts at around $810\, \mathrm{nm}$ in the idler output indicate a perfect separation of signal and idler photons}
	\label{Fig8}
\end{figure}

We perfomed a first characterization of the assembled module by pumping it with a cw-laser at $532\, \mathrm{nm}$ (Klastech) and analyzing the outputs with a spectrometer at visible wavelengths. We controlled the input power and polarization with a half-wave plate and a linear polarizer and coupled the light into the input PM-fiber of the HSPS module. The output fibers were directly connected to a spectrometer. Figure \ref{Fig8} shows the measured spectrum of the signal output port at three different temperatures measured with the integrated thermosensor. A comparison with the measured spectrum of the pure $\mathrm{Ti {:} LiNbO_3}$ waveguide (Figure \ref{Fig2}) indicates a working module with temperature wavelength tunability. The intensity drop for high temperatures can be explained by a small misalignment in the module due to different expansions of the components. Peaks from higher-order modes and Cerenkov radiation are clearly visible. We observe a perfect separation of signal and idler, since the spectrum of the idler output port shows no intensity counts at around $810\, \mathrm{nm}$ (see Figure \ref{Fig8}, green). Nevertheless, a small amount of pump photons is visible in both output ports, which indicates a insufficient pump filtering within the module. Thus we found that an additional filter is needed. We implemented this by fabricating a fiber-coupled filter for the signal output port. 
\subsection{Fiber-coupled narrow band filter}\label{subsec3}
For filtering the signal output port, we deposited a dielectric coating on a SM-fiber tip. We cutted the fiber near the fiber plug for insertion in the coating machine. We designed a dielectric coating consisting of $23$ alternating $\mathrm{TiO_2}$ and $\mathrm{SiO_2}$ layers. In order to ensure high quality, the layer thicknesses are measured optically in situ and the design is adjusted automatically if necessary. 
\begin{figure}[ht]%
	\centering
	\includegraphics[width=0.65\textwidth]{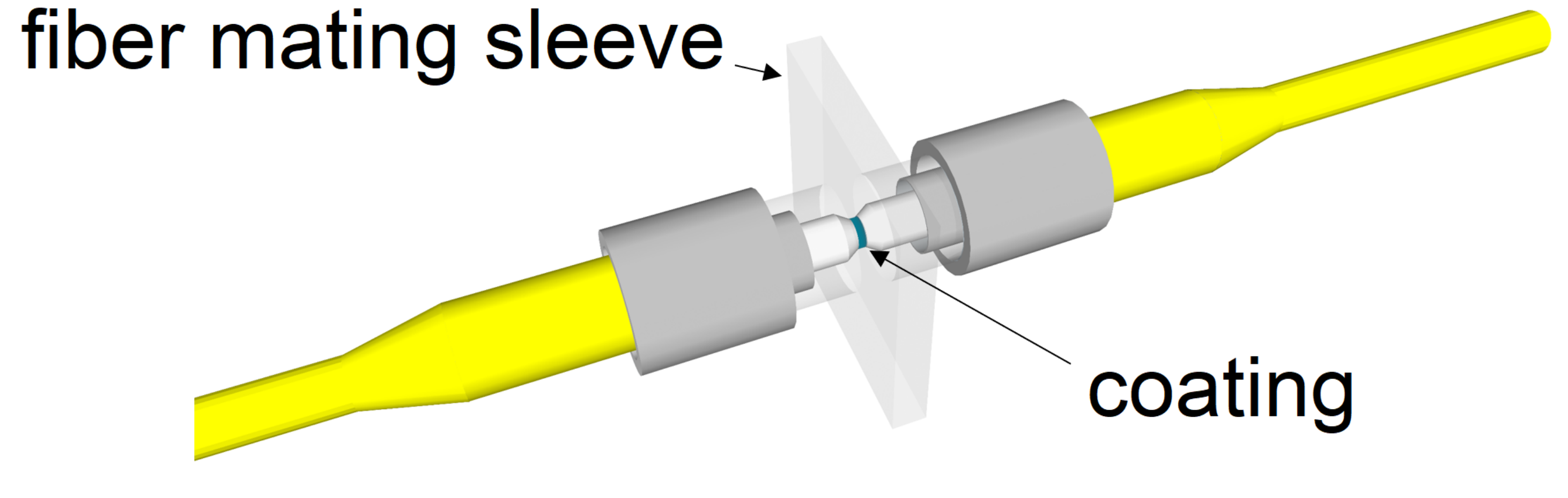}
	\caption{Sketch of the deposited coating between to SM-fiber tips for filtering the signal output port.}
	\label{Fig9}
\end{figure} \\
The transmission of a test glass coated simultaneously for a transmission measurement is shown in Figure \ref{Fig10} (black). We designed the coating to have a narrow band transmission at $810\, \mathrm{nm}$ not only for pump suppression but also for additional filtering of the Cerenkov radiation and possible luminescence of the used glue. We spliced the coated fiber tip to the original fiber and connected a second SM-fiber (see Figure \ref{Fig9}). The connected fiber tips are glued to the fiber mating sleeve for protection of the coating. The measured transmission through the assembled fiber-filter is shown in Figure \ref{Fig10} (red), with a transmission peak of $42\, \mathrm{\%}$ ($3.8 \, \mathrm{dB}$ losses) at $ 809.5\, \mathrm{nm}$. The spectrum of the HSPS module signal output port with connected fiber-filter is shown in Figure \ref{Fig11}. A comparison with the unfiltered module (Figure \ref{Fig8}) indicates the narrow band-pass filtering of the coated fiber.    
\begin{figure}[ht]
	\centering
	\begin{subfigure}[c]{0.49\linewidth}
		\centering
        \subcaption{}
		\includegraphics[width=\linewidth]{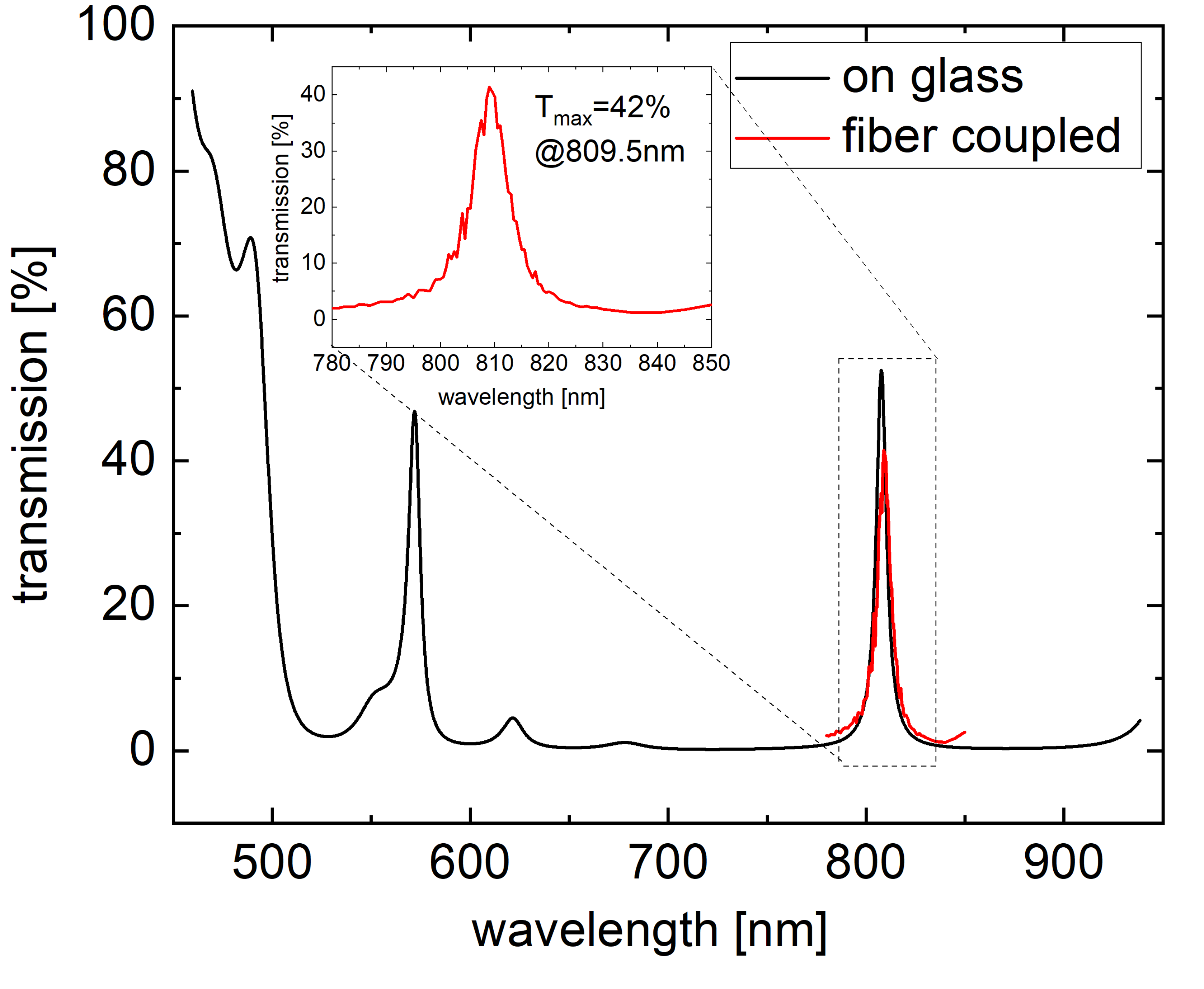}
		\label{Fig10}
    \end{subfigure}
	\begin{subfigure}[c]{0.49\linewidth}
		\centering
        \subcaption{}
		\includegraphics[width=\linewidth]{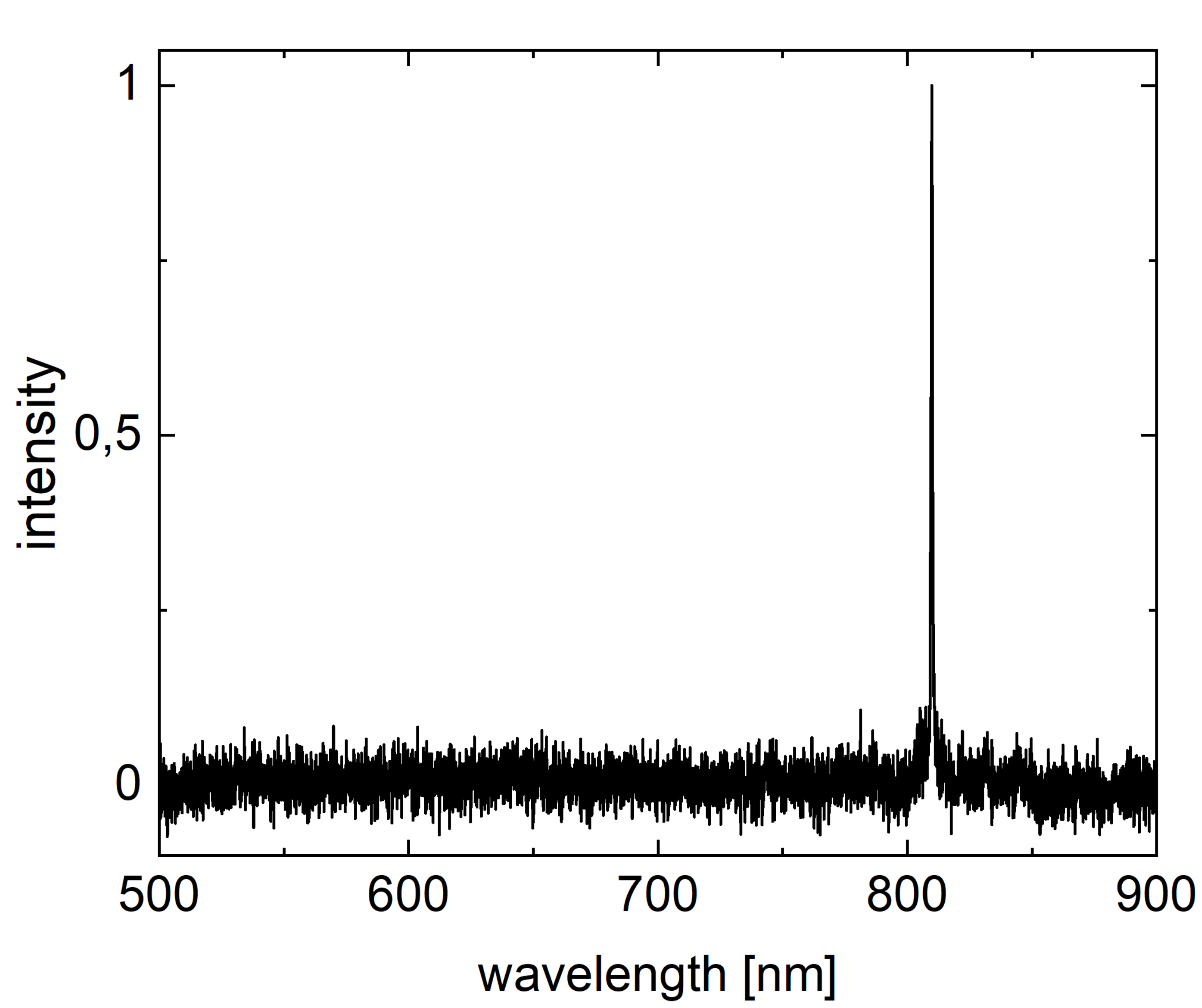}
		\label{Fig11}
    \end{subfigure}
    \caption{Transmission of the deposited coating, measured on a glass substrate (a, black) and after assembly of the fiber-filter (b, red); filtered spectrum of the HSPS module signal output port (b).}
\end{figure}
\section{Results and Discussion}\label{sec12}
\begin{figure}[ht]%
	\centering
	\includegraphics[width=1.0\textwidth]{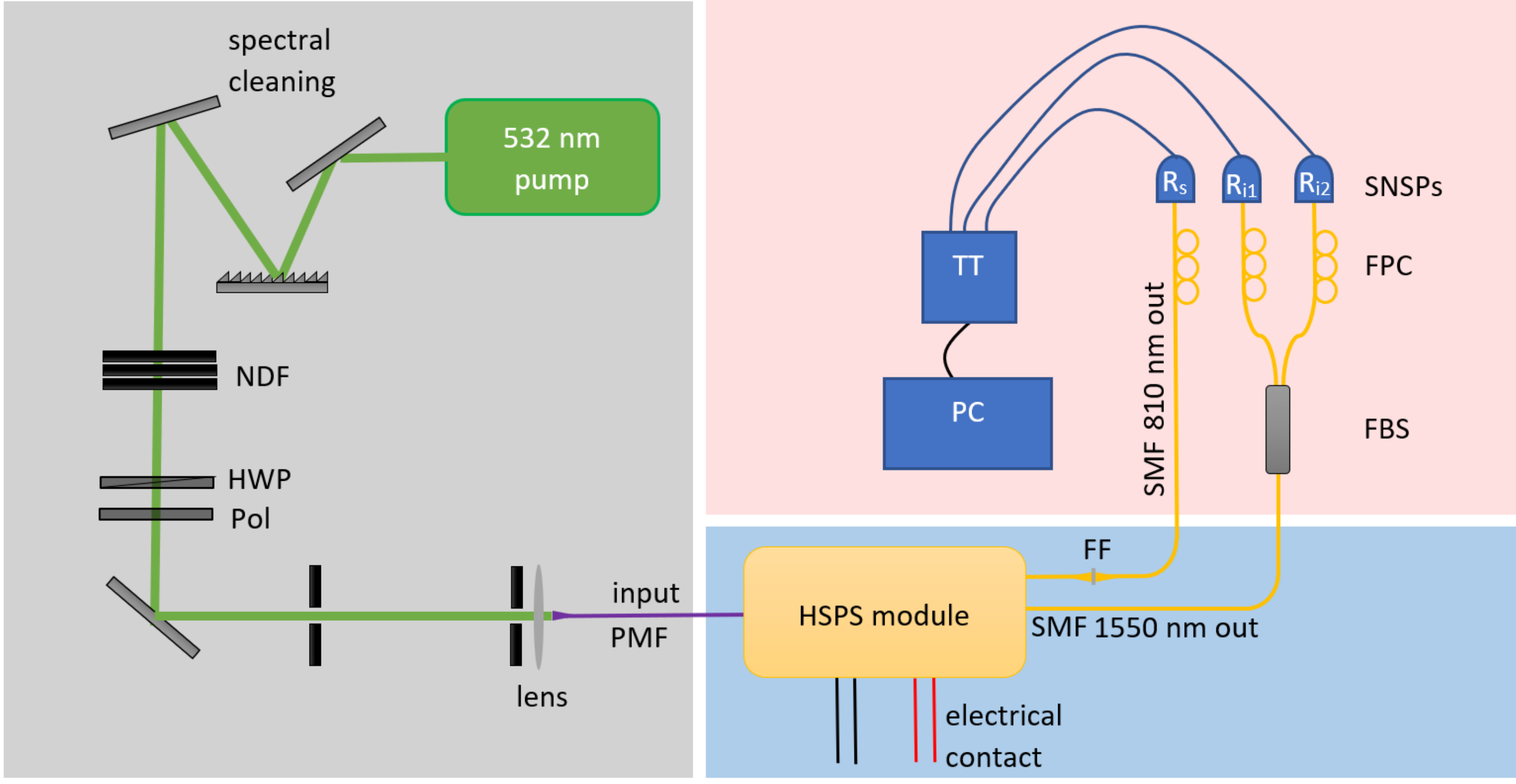}
	\caption{Experimental setup for conditioned measurements. NDF: neutral density filters; HWP: half-wave plate; Pol: linear polarizer; PMF: Polarization-maintaining fiber; SMF: single-mode fiber; FF: custom fiber-filter; FBS: fiber
		beamsplitter 50/50; FPC: fiber polarization controller; SNSPDs: superconducting nanowire single-photon detectors; TT: time-tagger}
	\label{Fig12}
\end{figure}

After we analyzed the spectral characteristic and temperature dependency of the module, we then investigated the single photon characteristics of the module by performing conditioned measurements with superconducting nanowire single-photon detectors. The experimental setup is shown in Figure \ref{Fig12}. We used a frequency-doubled pulsed pump source (Katana-05, Onefive GmbH Zürich) with a center wavelength of $532 \, \mathrm{nm}$ and a spectral width of $0.19 \, \mathrm{nm}$ to operate the modul. The Gaussian shaped pulses have a repetition rate of $10 \, \mathrm{MHz}$ with  pulses of $43 \, \mathrm{ps}$. To filter out background light, we used a blazed-grating after the laser. We adjusted the intensity and polarization by neutral density filters (NDF) and a combination of a half-wave plate (HWP) and a linear polarizer (Pol). The light was then coupled into a PM-fiber, which is connected to the input fiber of the module. We connected the filtered output of the signal to a SNSPD (Cryospot 4, Photon Spot California) with a detection efficiency of around $\eta_{Det,s}=65 \, \mathrm{\%}$ @ $810 \, \mathrm{nm}$ for click-detection (rate $R_s$). The output of the idler was coupled to a 50/50 fiber beamsplitter (FBS) and the photon rates ($R_{i1}$ and $R_{i1}$) are detected by two SNSPDs (detection efficiency of $\eta_{Det,i}=85 \, \mathrm{\%}$ @ $1550 \, \mathrm{nm}$). By splitting one of the outputs and detecting both rates individually, the generation of more than one photon pair can be measured. The threefold coincidence between all detectors indicates the generation of at least two photon pairs.   

\begin{figure}[ht]%
	\centering
	\includegraphics[width=0.75\textwidth]{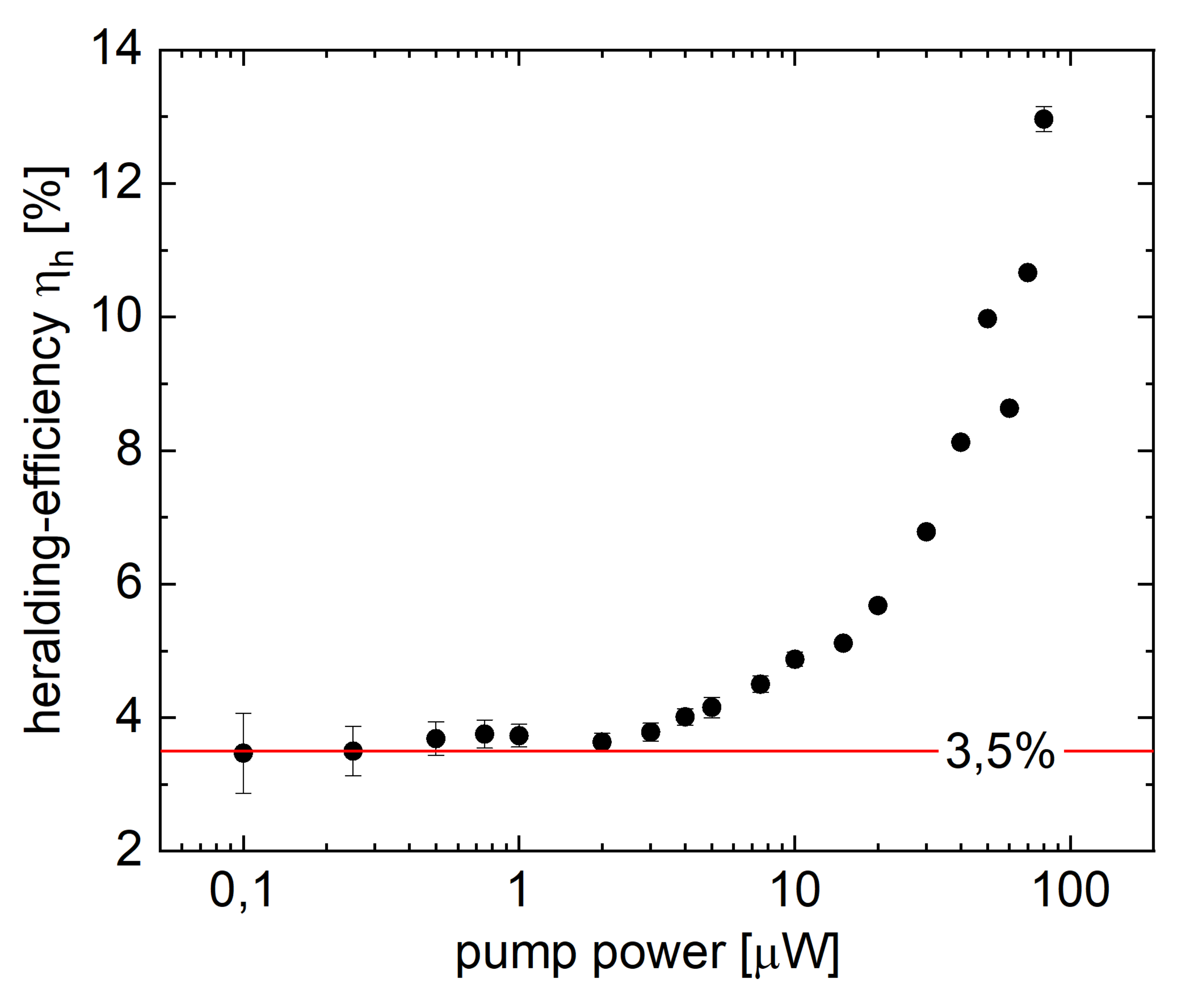}
	\caption{Heralding efficiency $\eta_h$ vs. pump power of the filtered HSPS module. For low pump powers the heralding efficiency approaches a vlaue of $\eta_h=3.5 \, \mathrm{\%}$.}
	\label{Fig13}
\end{figure}

\begin{figure}[ht]
	\centering
	\begin{subfigure}[c]{0.49\linewidth}
		\centering
        \subcaption{}
		\includegraphics[width=\linewidth]{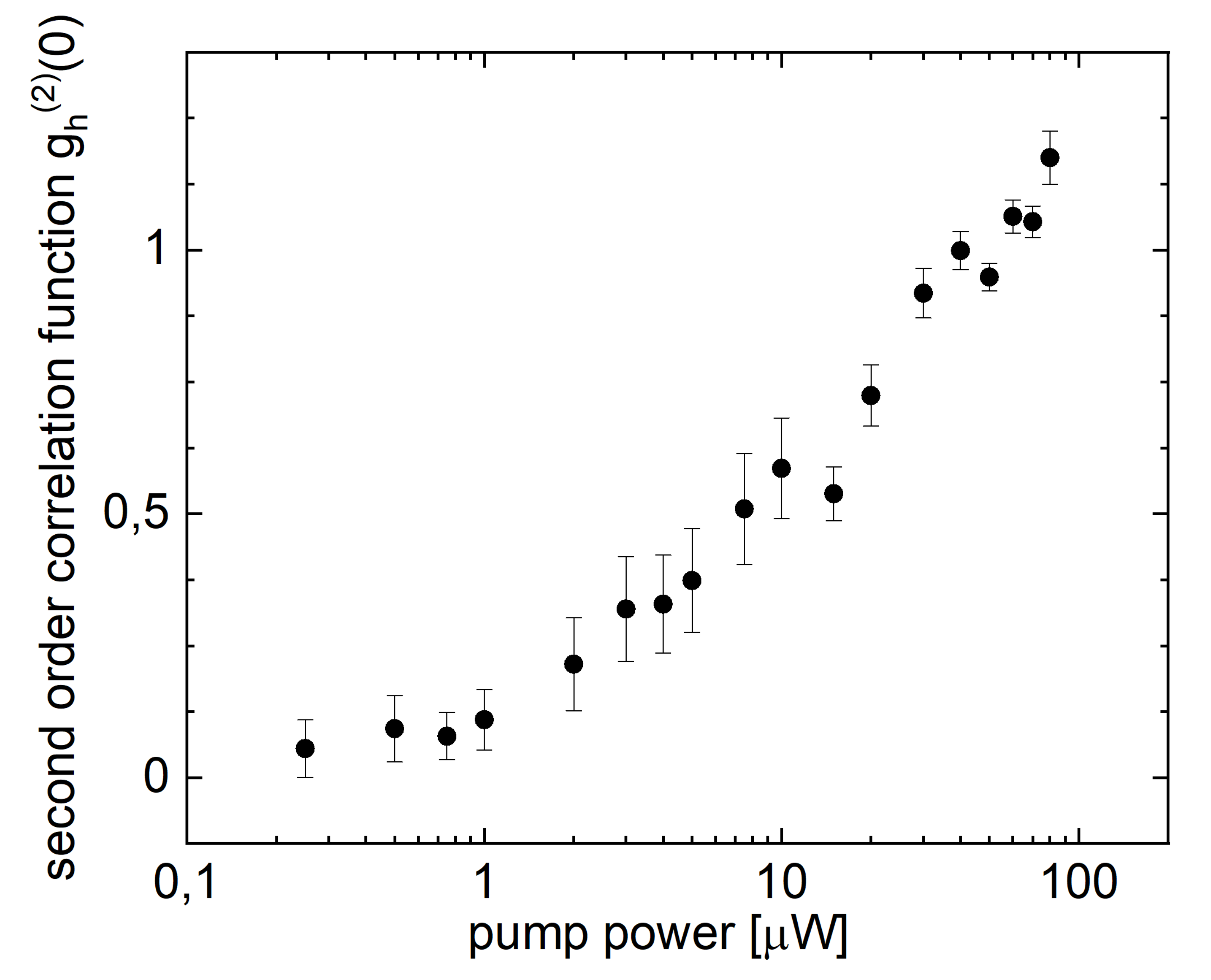}
		\label{Fig14}
    \end{subfigure}
	\begin{subfigure}[c]{0.49\linewidth}
		\centering
        \subcaption{}
		\includegraphics[width=\linewidth]{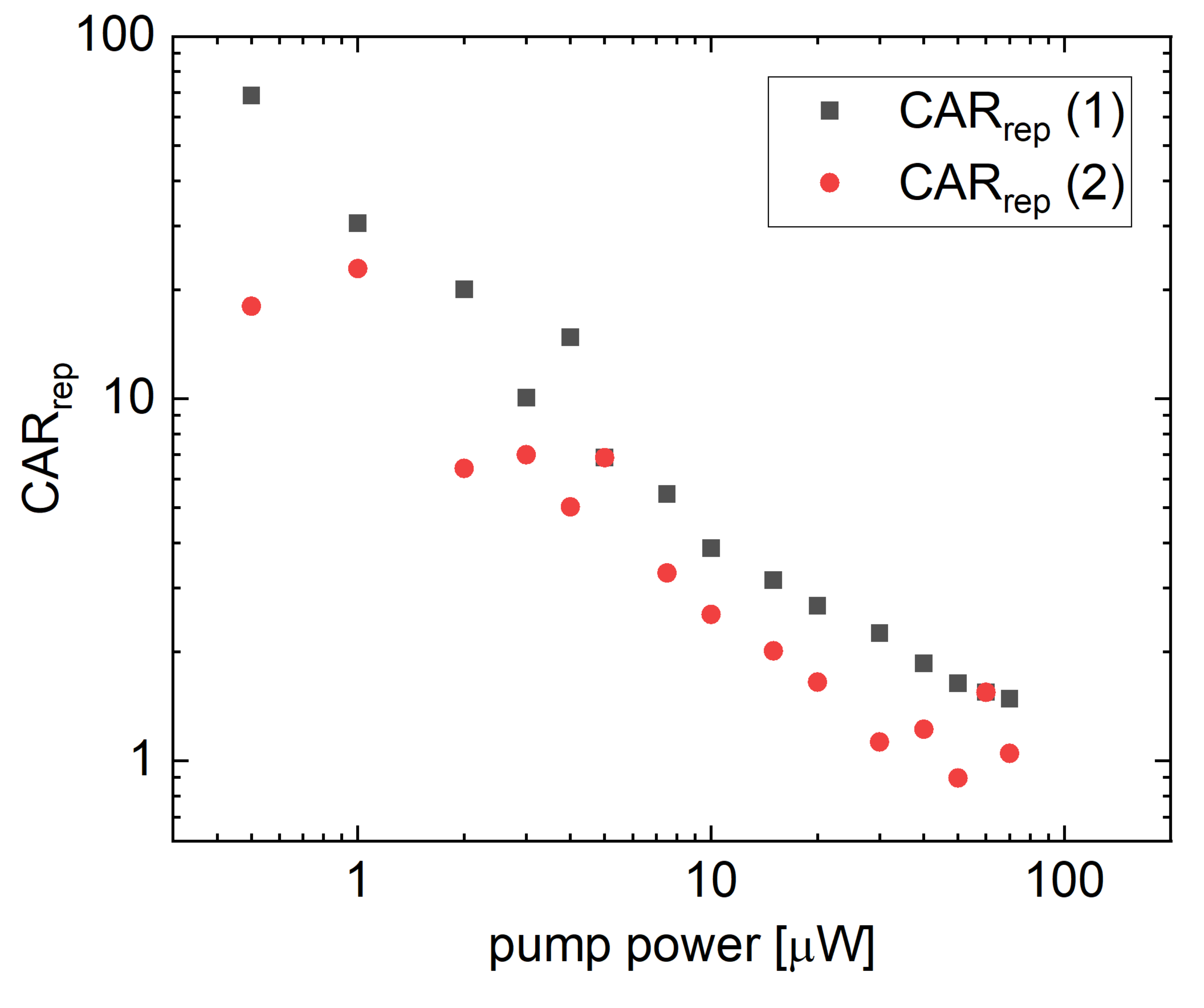}
		\label{Fig15}
    \end{subfigure}
    \caption{Heralded second-order auto-correlation function $g^2_h(0)$ vs. pump power (a). An increase for higher pump powers indicates the generation of higher-order photon components; time shifted coincidence-to-accidental ratio $CAR_{rep}(m)$ vs. pump power for time shifts of one and two pulse repetition times (b).}
\end{figure}

\newpage
We determined the single count rates and coincidences between the three used SNSPDs by a time-tagger (TT). To determine the single photon and heralding performance of the module, we evaluated the heralding-efficiency \cite{Krapick2013}  

\begin{equation}
	\eta_h=\frac{R_{s \wedge i1}+R_{s \wedge i2}}{R_s \cdot \eta_{Det,i}} ,
\end{equation}

where $R_{s \wedge i1}$ and $R_{s \wedge i2}$ describe the coincidences between signal and idler 1/idler 2 with a coincidence window of $5 \, \mathrm{ns}$. The heralding-efficiency $\eta_h$ describes the probability of a single photon in the idler output upon the detection event in the signal output. The power dependent heralding efficiency of the filtered module is shown in Figure \ref{Fig13}. The heralding efficiency increases with higher pump powers, due to an increase of higher order photon components, which is caused by the high brightness of SPDC waveguide sources. For low pump powers, we can obtain a constant heralding efficiency of around $\eta_h=3.5 \, \mathrm{\%}$, which is low compared to other works \cite{Signorini2020}. It is important to mention that our module is a hybrid integration, which is not the case with the comparable sources. Non-optimal coupling between the two material systems leads to additional losses that reduce the heralding efficiency. To estimate the total losses within the assembled module, we performed count rate measurements with different additional filter combinations. We measured the output count-rates of the module without any filters and after the insertion of the built fiber-coupled narrow band filter in the signal port and another narrow band filter in the idler port. Count-rates drop by around $90 \, \mathrm{\%}$ in both outputs after insertion of the filters. This indicates a background in the unfiltered module of around $90 \, \mathrm{\%}$, which should be mainly due to residual pump light. As described in chapter \ref{subsec3} the background can be completly reduced by inserting the narrow band filter. We estimate the transmission efficiency through the module for the signal path ($\eta_s$) and idler path ($\eta_i$) by using the measured countrates $R_s$, $R_i$ and $R_{s \wedge i}$ and the following equations

\begin{equation}
\begin{aligned}
	R_s&=\eta_s \cdot \eta_{filter,s} \cdot R_{SPDC} \\
	R_i&=\eta_i \cdot \eta_{filter,i} \cdot R_{SPDC} \\
	R_{s \wedge i}&=\eta_s \cdot \eta_{filter,s} \cdot \eta_i \cdot \eta_{filter,i} \cdot R_{SPDC} ,
\end{aligned} 
\end{equation}  

where $R_{SPDC}$ is the SPDC photon pair rate generated in the $\mathrm{Ti {:} LiNbO_3}$ waveguide and $\eta_{filter,s}$ and $\eta_{filter,i}$ are the transmission efficiencies of the additional filters. We calculated the transmission efficiencies through the module to be $\eta_s=1.5 \, \mathrm{\%}$ ($18\, \mathrm{dB}$ losses) and $\eta_i=5 \, \mathrm{\%}$ ($13\, \mathrm{dB}$ losses). A comparison to the ideal losses (see \ref{subsec2}) indicates a low coupling efficiency between different components in the module.  
\\
To evaluate the simultaneous generation of multiple photon pairs in the generated SPDC , we make use of the heralded second-order auto-correlation function $g^2_h(0)$, which is given by \cite{Krapick2013} 
\begin{equation}
	g^2_h(0)=4\frac{R_s \cdot R_{s \wedge i1 \wedge i2}}{R_{s \wedge i1}+R_{s \wedge i2}}
\end{equation}

for the given measurement scheme. $R_{s \wedge i1 \wedge i2}$ is the threefold coincidence between signal, idler 1 and idler 2, which is only present if multiple photon pairs  are generated. For a perfect SPDC prozess, the $g^2_h(0)$ should therefore vanish. Figure \ref{Fig14} shows the measured $g^2_h(0)$ of the filtered module for different pump powers.
For low pump powers $<1 \, \mathrm{\mu W}$ the second-order auto-correlation function reaches values of $g^2_h(0)<0.005$, which indicates an excellent single photon character and is comparable to a previous HSPS based on $\mathrm{Ti {:} LiNbO_3}$, developed and fabricated in our group \cite{Krapick2013}. For a further investigation of two photon pair generation and an analysis of possible photoluminescence within the module, we calculate a coincidence-to-accidental ratio $CAR_{rep}(m)$. The CAR is the ratio between  coincidences of photons in the signal and idler outputs for one input pulse $R_{s \wedge i}(\Delta \tau=0)$ and for coincidences with neighbouring pulses $R_{s \wedge i}(\Delta \tau=m\tau_{rep})$. We shift the relative time delay between signal and idler such that a pulse overlaps with a neighbouring pulse. $\tau_{rep}$ is the repetition time of the pulsed laser and $m$ is a natural number, which gives the number of skipped pulses . $CAR_{rep}(m)$ is given by \cite{Krapick2013}      
\begin{equation}
	CAR_{rep}(m)=\frac{R_{s \wedge i}(\Delta \tau=0)}{R_{s \wedge i}(\Delta \tau=m\tau_{rep})}   \ \ m \in \mathbb{N} .
\end{equation}

Regardless of the number of skipped pulses $m$, $CAR_{rep}(m)$ should be constant. Figure \ref{Fig15} shows the power dependent $CAR_{rep}(m)$ for shifts of one and two pulses. The absolute values are again comparable to other works \cite{Krapick2013}. As can be seen, a larger pulse shift does not changes the $CAR_{rep}(m)$ significantly, which underlines the excellent single photon character. A decrease for higher pump powers indicates contributions of non-correlated photons. Therefore, the HSPS has to be operated at pump powers $<1\, \mathrm{\mu W}$, which however leads to a low heralding efficiency due to the excess losses in the module discussed above.     

\section{Conclusion and outlook}\label{sec13}
We demonstrated the first chip-size ($(2 \times 1)\, \mathrm{cm^2}$) fully integrated fiber-coupled Heralded Single Photon Source module based on a hybrid integration of a $\mathrm{Ti {:} LiNbO_3}$ waveguide into a polymer board with integrated optical components. The module can be pumped at $532\, \mathrm{nm}$ via a input PM-fiber. A SPDC process in a low-loss periodically poled $\mathrm{Ti {:} LiNbO_3}$ waveguide creates signal and idler photons at $810\, \mathrm{nm}$ and $1550\, \mathrm{nm}$. The nonlinear crystal is coupled to a PolyBoard with integrated waveguides and filters for pump suppression and dichroic photon separation. A characterization showed an operating module with temperature tunability. Nevertheless, measurements of the output spectra showed residual pump light in the outputs. \\
Therefore, we fabricated a fiber-coupled narrow band pass filter, with a high transmission at $810 \, \mathrm{nm}$ by depositing a $23$ layer dielectric coating onto a SM-fiber tip.
We measured a heralding efficiency of $\eta_h=4.5\, \mathrm{\%}$ for the filtered module at low pulsed pump powers, which is comparatively low and can be explained by bad coupling efficiencies within the module. Calculations of a pulse dependent coincidence-to-accidental ratio and a heralded second-order correlation function of $g_h^{(2)}=0.05$ show a excellent single photon character. \\
Significantly improving the coupling in the module could greatly lower losses, which should result in a higher heralding efficiency. 
In addition, the hybrid integration of $\mathrm{Ti {:} LiNbO_3}$ into a PolyBoard opens up the opportunity for a future multi-channel source and a direct integration of silicon SPAD arrays for the detection of the signal photons on chip. The realization of a quantum- system-on-chip (QSoC) providing low cost, small size and robustness is therefore conceivable with a hybrid integration approach.  
\\ \\
\textbf{Acknowledgments.} This work was funded by the European Union’s Horizon
2020 Research and Innovation Programme through Quantum-Flagship Project
UNIQORN under Grant 820474.
\\ \\
\textbf{Disclosures.} The authors declare no conflicts of interest.

\bibliography{HSPS}

\begin{thebibliography}{10}
\newcommand{\enquote}[1]{``#1''}

\bibitem{Obrien2008}
J.~L. O{\textquotesingle}Brien, \enquote{Optical quantum computing,}
  {\protect\JournalTitle{Science}} \textbf{318}, 1567--1570 (2007).

\bibitem{Krenn2016}
Mehul, S.~Thomas, U.~Rupert, Z.~A.~K. Mario, and Malik, \emph{Quantum
  Communication with Photons} (Springer International Publishing, 2016).

\bibitem{Arakawa2020}
Y.~Arakawa and M.~J. Holmes, \enquote{Progress in quantum-dot single photon
  sources for quantum information technologies: A broad spectrum overview,}
  {\protect\JournalTitle{Applied Physics Reviews}} \textbf{7} (2020).

\bibitem{Senellart2021}
P.~Senellart, \enquote{Semiconductor single-photon sources: progresses and
  applications,} {\protect\JournalTitle{Photoniques}} pp. 40--43 (2021).

\bibitem{Senellart2017}
P.~Senellart, G.~Solomon, and A.~White, \enquote{High-performance semiconductor
  quantum-dot single-photon sources,} {\protect\JournalTitle{Nature
  Nanotechnology}} \textbf{12}, 1026--1039 (2017).

\bibitem{Krapick2013}
S.~Krapick, H.~Herrmann, V.~Quiring, B.~Brecht, H.~Suche, and C.~Silberhorn,
  \enquote{An efficient integrated two-color source for heralded single
  photons,} {\protect\JournalTitle{New Journal of Physics}} \textbf{15} (2013).

\bibitem{Shukhin2020}
A.~A. Shukhin, J.~Keloth, K.~Hakuta, and A.~A. Kalachev, \enquote{Heralded
  single-photon and correlated-photon-pair generation via spontaneous four-wave
  mixing in tapered optical fibers,} {\protect\JournalTitle{Physical Review A}}
  \textbf{101} (2020).

\bibitem{Paesani2020}
S.~Paesani, M.~Borghi, S.~Signorini, A.~Maïnos, L.~Pavesi, and A.~Laing,
  \enquote{Near-ideal spontaneous photon sources in silicon quantum photonics,}
  {\protect\JournalTitle{Nature Communications}} \textbf{11} (2020).

\bibitem{Bock2016}
M.~Bock, A.~Lenhard, C.~Chunnilall, and C.~Becher, \enquote{Highly efficient
  heralded single-photon source for telecom wavelengths based on a ppln
  waveguide,} {\protect\JournalTitle{Optics Express}} \textbf{24}, 23992
  (2016).

\bibitem{Tanzilli2001}
S.~Tanzilli, H.~Riedmatten, H.~Tittel, H.~Zbinden, P.~Baldi, M.~De~Micheli,
  D.~Ostrowsky, and N.~Gisin, \enquote{Highly efficient photon-pair source
  using a periodically poled lithium niobate waveguide,}
  {\protect\JournalTitle{Electronics Letters}} \textbf{37}, 26 -- 28 (2001).

\bibitem{Fujii2022}
G.~Fujii, N.~Namekata, M.~Motoya, S.~Kurimura, S.~Inoue, N.~Gisin, G.~Ribordy,
  W.~Tittel, H.~Zbinden, J.~W. Pan, D.~Bouwmeester, H.~Weinfurter, and
  A.~Zeilinger, \enquote{Quantum optics; (190.4410) nonlinear optics,}
  {\protect\JournalTitle{Rev. Mod. Phys}} \textbf{74}, 145--195 (2002).

\bibitem{Meany2014}
T.~Meany, L.~A. Ngah, M.~J. Collins, A.~S. Clark, R.~J. Williams, B.~J.
  Eggleton, M.~J. Steel, M.~J. Withford, O.~Alibart, and S.~Tanzilli,
  \enquote{Hybrid photonic circuit for multiplexed heralded single photons,}
  {\protect\JournalTitle{Laser and Photonics Reviews}} \textbf{8} (2014).

\bibitem{Montaut2017}
N.~Montaut, L.~Sansoni, E.~Meyer-Scott, R.~Ricken, V.~Quiring, H.~Herrmann, and
  C.~Silberhorn, \enquote{High-efficiency plug-and-play source of heralded
  single photons,} {\protect\JournalTitle{Physical Review Applied}} \textbf{8}
  (2017).

\bibitem{Schlehahn2018}
A.~Schlehahn, S.~Fischbach, R.~Schmidt, A.~Kaganskiy, A.~Strittmatter, S.~Rodt,
  T.~Heindel, and S.~Reitzenstein, \enquote{A stand-alone fiber-coupled
  single-photon source,} {\protect\JournalTitle{Scientific Reports}} \textbf{8}
  (2018).

\bibitem{Vergyris2022}
P.~Vergyris, \enquote{Harnessing the power of single photons,}
  {\protect\JournalTitle{PhotonicsViews}} \textbf{19}, 26--29 (2022).

\bibitem{Aurea2023}
{Aurea Technology}, \enquote{{Twin Photon Source at telecom wavelengths},}
  https://www.aureatechnology.com/en/products/twin-photons-source.html (2023).

\bibitem{Keil2012}
N.~Keil, C.~Zawadzki, D.~de~Felipe, Z.~Zhang, and N.~Grote, \enquote{Polymer
  based platform for hybrid photonic integration,}  (2012).

\bibitem{Kleinert2017}
M.~Kleinert, D.~de~Felipe, C.~Zawadzki, W.~Brinker, J.~H. Choi, P.~Reinke,
  M.~Happach, S.~Nellen, M.~Möhrle, H.-G. Bach, N.~Keil, and M.~Schell,
  \enquote{Photonic integrated devices and functions on hybrid polymer
  platform,} {\protect\JournalTitle{Physics and Simulation of Optoelectronic
  Devices XXV}} \textbf{10098}, 100981A (2017).

\bibitem{Weis1985}
R.~S. Weis and T.~K. Gaylord, \enquote{Lithium niobate: Summary of physical
  properties and crystal structure,} {\protect\JournalTitle{Appl. Phys. A}}
  \textbf{37}, 191--203 (1985).

\bibitem{Minakata1978}
M.~Minakata, S.~Saito, M.~Shibata, and S.~Miyazawa, \enquote{Precise
  determination of refractive-index changes in ti-diffused linbo 3 optical
  waveguides,} {\protect\JournalTitle{Journal of Applied Physics}} \textbf{49},
  4677--4682 (1978).

\bibitem{Fukuma1978}
M.~Fukuma and H.~Iwasaki, \enquote{A study on titanium diffusion into linb03
  waveguides by electron probe analysis and x-ray diffraction methods,}
  {\protect\JournalTitle{JOURNAL OF MATERIALS SCIENCE}} \textbf{13}, 523--533
  (1978).

\bibitem{Zhang2008}
Y.~Zhang, Z.~D. Gao, Z.~Qi, S.~N. Zhu, and N.~B. Ming, \enquote{Nonlinear
  čerenkov radiation in nonlinear photonic crystal waveguides,}
  {\protect\JournalTitle{Physical Review Letters}} \textbf{100} (2008).

\bibitem{Mondain2020}
F.~Mondain, F.~Brunel, X.~Hua, E.~Gouzien, A.~Zavatta, T.~Lunghi, F.~Doutre,
  M.~P.~D. Micheli, S.~Tanzilli, and V.~D’Auria, \enquote{Photorefractive
  effect in linbo 3 -based integrated-optical circuits for continuous variable
  experiments,} {\protect\JournalTitle{Optics Express}} \textbf{28}, 23176
  (2020).

\bibitem{Kleinert2019}
M.~Kleinert, M.~Nuck, H.~Conradi, D.~de~Felipe, M.~Kresse, W.~Brinker,
  C.~Zawadzki, N.~Keil, and M.~Schell, \enquote{A platform approach towards
  hybrid photonic integration and assembly for communications, sensing, and
  quantum technologies based on a polymer waveguide technology,}  (2019), pp.
  25--30.

\bibitem{Signorini2020}
S.~Signorini and L.~Pavesi, \enquote{On-chip heralded single photon sources,}
  {\protect\JournalTitle{AVS Quantum Science}} \textbf{2} (2020).

\end{thebibliography}






\end{document}